\documentclass[fleqn,10pt]{wlscirep}
\usepackage[utf8]{inputenc}
\usepackage[T1]{fontenc}
\usepackage{graphicx}
\usepackage{float}	
\usepackage{ucs}
\usepackage{amsmath}  
\usepackage{amsfonts}
\usepackage{amssymb}
\usepackage{booktabs}
\usepackage{textcomp}
\usepackage{gensymb}
\usepackage{enumitem}
\usepackage{multirow}
\usepackage{lineno}

\title{MgO surface lattice phonons observation during Interstellar ice transition}

\author[1,2]{A. Chavarr\'{i}a--Sibaja}
\author[1,2]{S. Mar\'{i}n--Sosa}
\author[2,3]{E. Bolaños--Jiménez}
\author[4]{M. Hernández--Calder\'{o}n}
\author[1,2,5,6*]{O.A. Herrera--Sancho}
\affil[1]{Escuela de Física, Universidad de Costa Rica, 2060 San Pedro, San José, Costa Rica}
\affil[2]{Centro de Investigación en Ciencia e Ingeniería de Materiales, Universidad de Costa Rica, 2060 San Pedro, San José, Costa Rica}
\affil[3]{Escuela de Ingeniería Química, Universidad de Costa Rica, 2060 San Pedro, San Jos\'e, Costa Rica.}
\affil[4]{Escuela de Ciencia e Ingenier\'{i}a de Materiales, Instituto Tecnol\'{o}gico de Costa Rica, 30101, Cartago, Costa Rica}
\affil[5]{Centro de Investigación en Ciencias Atómicas Nucleares y Moleculares, Universidad de Costa Rica, 2060 San Pedro, San José, Costa Rica}
\affil[6]{Instituto de Investigaciones en Arte, Universidad de Costa Rica, 2060 San Pedro, San José, Costa Rica}

\affil[*]{oscar.herrerasancho@ucr.ac.cr}

\begin{abstract}

Relevant information on the origins of the solar system and the early evolution of life itself can be derive from systematic and controlled exploration of water ice here on Earth. Therefore, over the last decades, a huge effort on experimental methodologies has been made to study the multiple crystal ice phases, which are observed outside our home--gravitational--potential. By employing (100)--oriented MgO lattice surface as a microcantilever sensor, we conducted the first ever study on the dynamics of the Structural Phase Transition at 185\,K in water ice by means of coherent elastic scattering of electron diffraction. We estimate the amount of phonons caused by this transition applying precise quantum computing key tools, and resulting in a maximum value of 1.23\,$\pm$\,0.02. Further applications of our microcantilever sensor were assessed using unambiguous mapping of the surface stress induced by the \textit{c}(4$\times$2) $\longrightarrow$ \textit{p}(3$\times$2) Structural Phase Transition of the interstellar ice formulated on the Williamsom--Hall model. This development paves the way and thus establishes an efficient characterization tool of the surface mechanical strains of materials with potential applications arising from interstellar ice inclusive glaciers to the wide spectrum of solid--state physics. 

\end{abstract}

\begin{document}

\flushbottom
\maketitle
%
%
\thispagestyle{empty}


\section*{Introduction}
\label{Introduction}
\textit{``The hope is that if we do understand the ice crystal, we shall ultimately understand the glacier''}, said the famous Nobel Prize Richard Feynman in one of his lectures titled \textit{The Character of Physical Law}~\cite{bartels2012ice}. With this phrase, Feynman exemplifies the implications of reaching a deep understanding of the physical and chemical ice crystal properties. These properties are related to its molecular morphology that depends on the order of the hydrogen--bonds configuration~\cite{loerting2020open}. Hence, the ice presents stable configurations like cubic (I$_{c}$) and hexagonal (I$_{h}$) structures, meta--stable amorphous states with high and low density structures (I$_{a}$h and I$_{a}$l), and restricted amorphous structural states (I$_{a}$r)~\cite{jenniskens1994structural,jenniskens1995high}. There is an increasing body of literature that suggests a relation between the Structural Phase Transition (SPT) of ice with photolysis and the formation of ionic radicals in comets and other astrophysical media; making it relevant in the understanding of the origins of the earlier solar system and life on Earth~\cite{kouchi1990amorphization,stevenson1999controlling,bernstein2002racemic,meinert2016ribose}. This underscores the need to understand the crystallization and SPT's to unravel the still hidden secrets of the ice--related phenomena.

However, water ice is a familiar well--studied substance, the same is dissimilar for its counterpart, interstellar ice. This indicates a need to understand the various perceptions of interstellar ice, which has motivated the development of methods for the characterization of SPT of the ice structures and related interactions~\cite{mishima1984melting,jenniskens1996crystallization,li1997inelastic,16doc-xu1997structure,jenniskens1998amorphous,tulk2019absence}. Notably, the reactivity of water towards a metal oxide surface can be a fundamental tool in the understanding of these interactions~\cite{maboudian1997critical,marcus1998surface}. Among the oxides, magnesium oxide (MgO) has been the subject of many studies due to its high ionic character and simple crystal structure~\cite{23doc-wlodarczyk2011structures}. Its abilities on catalysis, adsorption--desorption process, and its use as a lattice template for selective growth of thin films have been studied for different purposes~\cite{davis1991aromatization,10doc-halim2004surface,23doc-wlodarczyk2011structures,16doc-xu1997structure,kim2002dissociation,song2016strain}. MgO surface is an ideal platform for water interaction studies due to the formation of free OH group (OH$_{f}$) and surface OH groups (OH$_{s}$) that give a vibrational feature to observe the response of MgO via the interplay of the interestellar ice SPT. While some researchers have been carried out groundwork using MgO surface lattice as a quantum--ideal--terrace to determine spin lifetime~\cite{Paul2017}, magnetic dipolar field~\cite{Choi2017}, and giant magnetoresistance~\cite{Yuasa2004}; no studies have been found making use of this crystal lattice periodicity to exploit it as a microcantilever sensor to map intrinsic mechanical properties using electron diffraction. In addition, little is known about key requirements to use the well--stablished coherent interaction in quantum optics to approximate, for instance, time evolution of phonons in regular arrays as MgO surfaces.

The laboratory experiment described here were designed to research on the variation of the structural stress related to the absorption--desorption of ice monolayers in the MgO surface lattice, see Fig.~\ref{fig:Artistic}. Therefore, to the experiments described here, we induced the SPT between the interstellar ice structures with structural symmetries $c$(4$\times$2) and $p$(3$\times$2) at 185\,K~\cite{23doc-wlodarczyk2011structures} by the application of a precise heating rate that allows great process control. While quantum controlling this SPT, we also investigate the temporal transfer of phonons towards the crystal lattice applying an alternative method based on the Williamson--Hall stress model. We show that, by use of Low Energy Electron Diffraction, it is possible to obtain fundamental information of the strain, phonon number, and sidebands produced by the phase transition of interstellar ice over the microcantilever MgO crystal surface.


\subsection*{Dynamics of the interstellar ice around its Structural Phase Transition at 185\,K}

\begin{figure}[ht]
\includegraphics[width=1.0\textwidth]{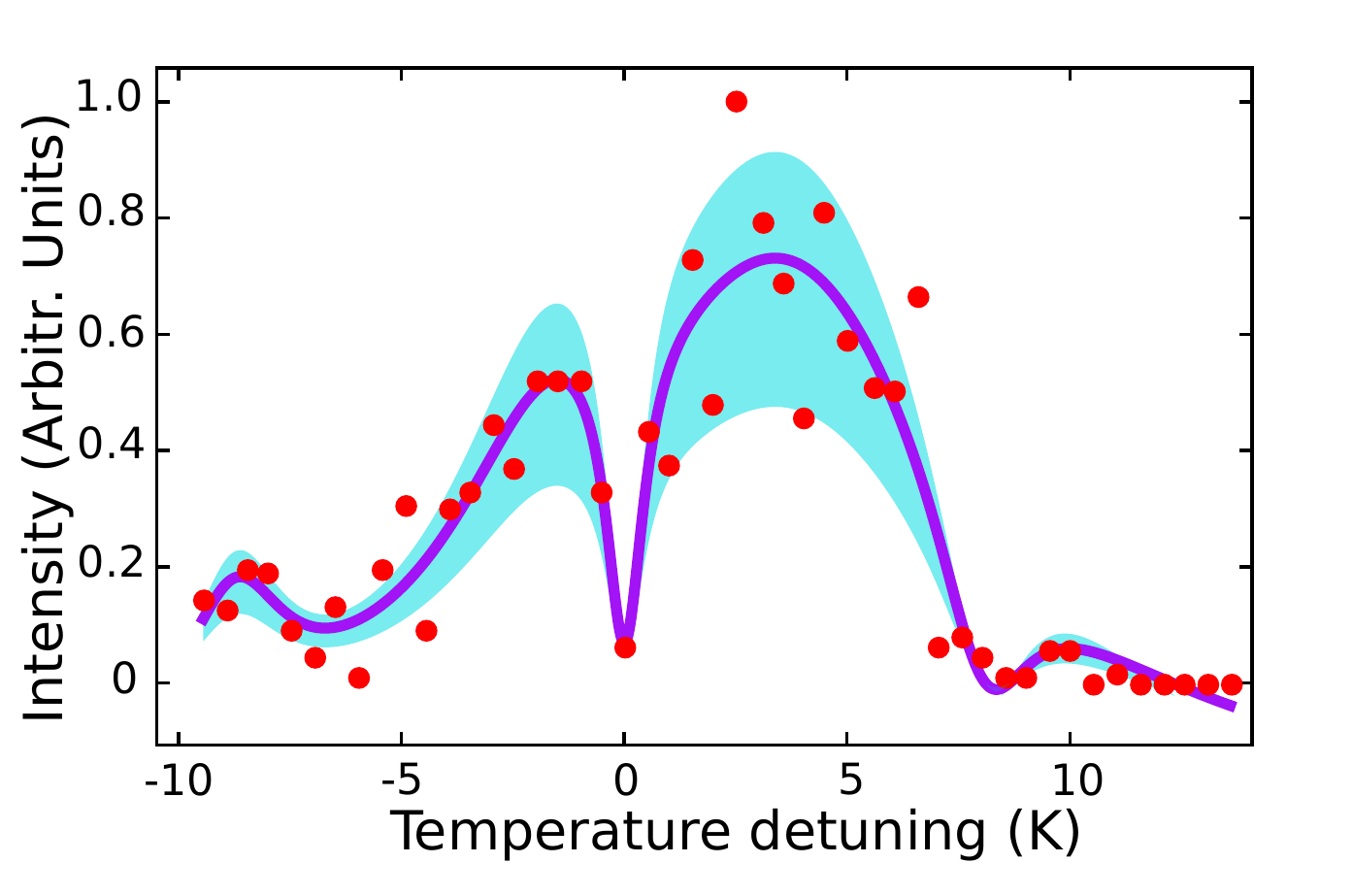}
\caption{\label{fig:IvsT} \textbf{Structural Phase Transition at 185\,K in water ice}. Here, the experimental heating rate corresponding to a well--controlled external--perturbation temperature of 0.55\,mK/s was used. Each experimental data point (red colour) corresponds to ten acquisition values of the intensity from a single diffracted spot (10). The energy of the incident electrons is 311\,eV. Clearly is seen the red and blue sideband features at approximately -9\,K and +10\,K, respectively. The \textit{c}(4$\times$2) $\longleftrightarrow$ \textit{p}(3$\times$2) transition of the ice at 185\,K is used as a reference temperature point (Temperature detuning equals to 0\,K). The purple curve corresponds to a Lorentzian fit to the data points. The cyan curve is the standard deviation of a Lorentzian fit to the data. See text for detailed information.}

\label{fig:phasetransition}
\end{figure}

To understand the dynamics of the interstellar ice over which an Structural Phase Transition (SPT) can temporarily progress, we architecturally grew water monolayers on top of the (100)--oriented MgO substrates. We investigated the time evolution of the SPT by monitoring the intensity of the diffraction spots from the Low--Energy Electron Diffraction (LEED) patterns along with full control and discretization of an external perturbation energy. 
All of our intensity measurements were performed on the spot (10) of the LEED diffraction pattern.
Figure~\ref{fig:phasetransition} provides experimental evidence obtained from the SPT in ice at about 185\,K by the fact of using the MgO surface as an unique sensor. Here, we observed, as expected, that when the temperature of the stabilized system is closed to the SPT, the diffraction spots become more intense~\cite{jenniskens1994structural} indicating the transfer of momentum from the solid water monolayers to the MgO surface. Importantly, the dropping of the diffraction intensity signal at 185\,K, which we correlate exactly with the SPT, corresponds to a lower probability that the eigenfunctions of the scattered electrons from the crystal structure contribute coherently in phase; suggesting an abrupt change in the sensor surface. The most interesting finding was to resolve the characteristic phonon distribution shown in Fig.~\ref{fig:phasetransition}, which was only feasible to obtain when the heating rate was reduced to approximately 3.0\,mK/s or below. In contrast to these findings, evidence of this discovery was not detected in previous studies since the heating rates used were around an order of magnitude higher~\cite{jenniskens1994structural,23doc-wlodarczyk2011structures}. This study has demonstrated for the first time that using controllable external perturbation temperature is highly desirable to explore the events that occur during SPT in interstellar ice. Furthermore, our results are consistent with those of other studies and suggest the existence of two stable surface structures: the low--temperature phase with \textit{c}(4$\times$2) symmetry before 185\,K and high--temperature phase with \textit{p}(3$\times$2) symmetry above 185\,K, containing about ten water molecules and six water moleculesper unit cell, respectively~\cite{23doc-wlodarczyk2011structures}. It is important to notice that as it can be observed in Fig.~\ref{fig:phasetransition}, increasing the temperature above 200\,K results in the complete desorption of water monolayers from the MgO (100) surface. Particularly, the presence of impurities, the deposition rate, style of deposition, heating rate, and annealing time have been identified as major contributing factors~\cite{jenniskens1994structural,Hagen1981} for the SPT to occur. To facilitate a comparison within experiments and reproducibility in our measurements, more than 100 experiments were carried out under circumstances in which the same experimental conditions (deposition rate and annealing time) for the 1.1\,mK/s and 0.55\,mK/s heating rates were employed. These experimental conditions contribute to maintaining constant the coverage factor and an amount of about two water monolayers on the MgO surface between the experiments.

We now briefly consider possible applications such as quantum computing and quantum information storage, focussing on the low--temperature and high--temperature SPT in the water monolayers. This indicates that the \textit{c}(4$\times$2) $\longleftrightarrow$ \textit{p}(3$\times$2) transition could be treated as a coherent Structural Phase Transition in order to enable abundant room for further progress in determining physical functionality and chemical reactivity in solids~\cite{Horstmann2020}. To confirm whether the SPT in interstellar ice exhibits coherent transition properties, we performed experiments where sinusoidal signatures of quantum coherence between the \textit{c}(4$\times$2) $\longleftrightarrow$ \textit{p}(3$\times$2) transition were temporarily searched by carefully switching the temperatures among these two states. Our data shows that there were no significant differences in the intensity of the diffraction spots in the \textit{c}(4$\times$2) $\longleftrightarrow$ \textit{p}(3$\times$2) transition in the time scale from seconds down to milliseconds. Future studies on the current topic are therefore recommended in order to monitor the structural transformation, for instance, by ultrafast Low--Energy Electron Diffraction~\cite{Horstmann2020}.

We now turn to the characteristic shape of our findings in Fig.~\ref{fig:phasetransition}, which resembles the phonon spectrum reported in previous studies for MgO lattice~\cite{Toyozawa1970,Sangster1970}. Therefore, here we addressed this phonon spectrum by tuning precisely the temperature of the system to enable a controlled SPT accompanied by the transfer of the phonon to the surface MgO sensor. We were able to identify the asymmetric sidebands (higher energy corresponds to diffraction spots with higher intensity) caused by impurities in the MgO crystal lattice as a consequence of the SPT local perturbation within the lattice that is directly mapped on our sensor. Importantly, we show for the first time that the SPT in interstellar ice could be monitored by means of (100)--oriented MgO crystal lattices. To our knowledge, this versatile platform detecting mechanical changes via electron diffraction has not been explicitly examined before.

\begin{figure}[ht]
\includegraphics[width=1.0\textwidth]{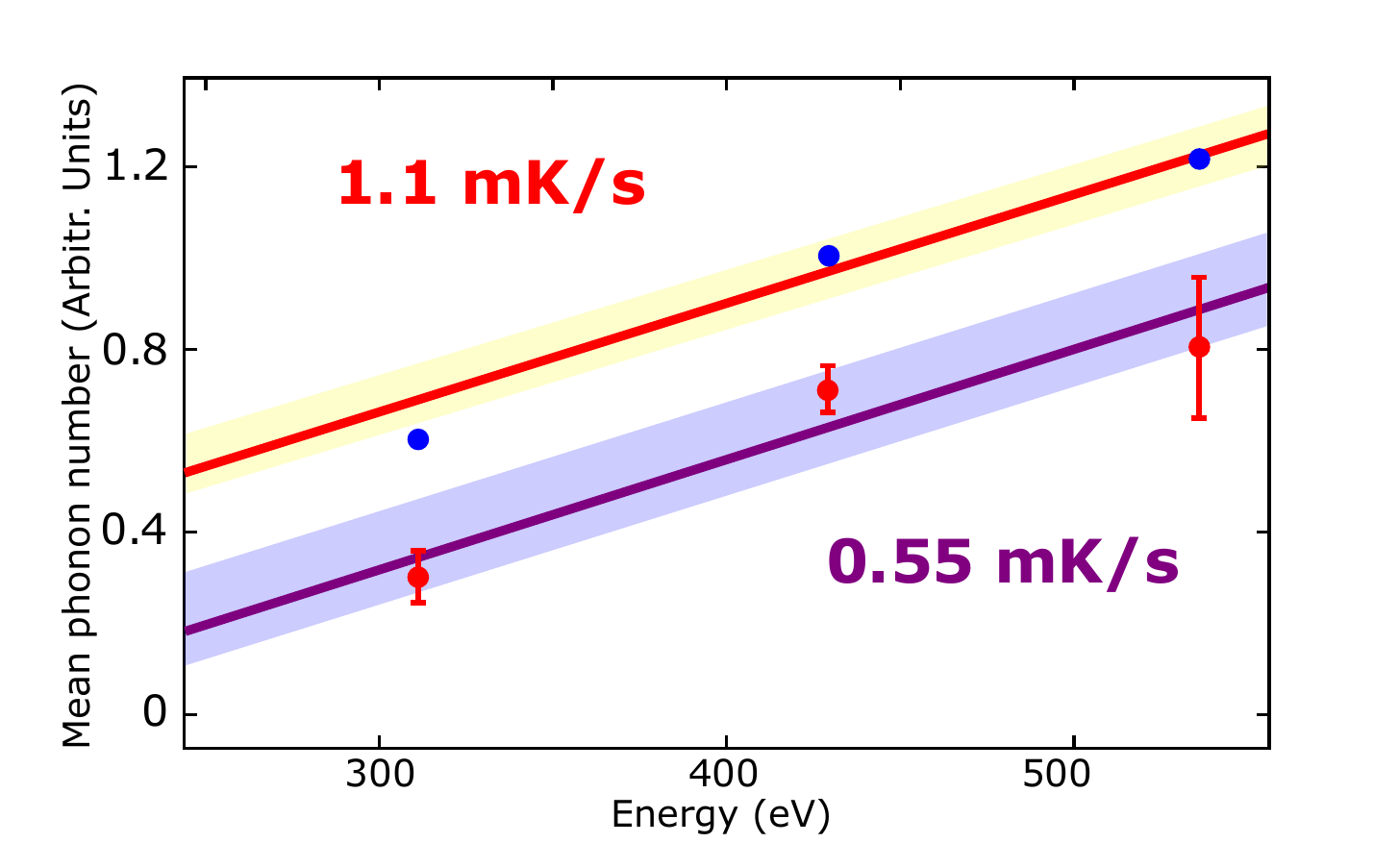}
\caption{\label{fig:PHvsE} \textbf{Penetration depth evolution of mean phonon number in the Structural Phase Transition at 185\,K.} Each data point represents the number of vibration modes of the MgO surface lattice according to the heating rate of 0.55\,mK/s (red dots) and 1.1\,mK/s (blue dots). Each measured value is equal to the mean over around 50 measurements and the error bars derive from the standard deviation. The violet and yellow shaded areas are the standard deviations of the linear fits, represented by the purple and red linear fits of the 0.55\,mK/s and 1.1\,mK/s heating rates, respectively. The diffraction spots below 311\,eV and above 536\,eV are unstable and ill--defined, therefore they were not considered for this analysis. The mean phonon number $\bar n$ in the MgO surface is inferred from the motional sidebands: amplitude ratio of the blue sideband over the red sideband, see Fig.~\ref{fig:phasetransition}). }
\label{fig:phononnumber}
\end{figure}

Next, we investigated how heating rates below the 3\,mK/s threshold would affect the dynamics of the phonon transferred to the MgO surface sensor. Figure~\ref{fig:phononnumber} shows the mean phonon number of two data sets for the following heating rates: 1.1\,mK/ and 0.55\,mK/s as a function of the energy of the incident electrons. In a similar manner to that used in the interaction of a laser with an ion trapped in a periodic potential~\cite{Leibfried2003}, we approximate the dynamics of the phonon in the MgO crystal by means of the following unperturbed Hamiltonian: $H_{0}=\frac{\hbar\omega_{0}}{2}\sigma_{z}+\hbar\omega_{t}(a^{\dagger}+1/2)=H_{a}+\hbar\omega_{t}(a^{\dagger}+1/2)$; where $\omega_{0}$ corresponds to the bare SPT transition energy, $\hbar$ is equal to the reduced Planck's constant ($h$) divided by 2$\pi$, $H_{a}$ is the bare two--level SPT transition energy diference between the two eigenstates (\textit{c}(4$\times$2) and \textit{p}(3$\times$2)) of the Pauli operator $\sigma_{z}$, $\omega_{t}$ indicates the phonon oscillation frequency in the trapping potential of the lattice, and $a^{\dagger}$ represents the creation operator. Previous studies have also examined the effect of motional sidebands produced by a macroscopic mechanical oscillator in order to infer the mean phonon number of the system~\cite{Underwood2015}. As a consequence of this approximation, the lattice’s mean phonon number $\bar n$ observed in the MgO surface is inferred only from the amplitude ratio~\cite{Weinstein2014} of the blue sideband (about +10\,K from the SPT in Fig.~\ref{fig:phasetransition}) and the red sideband (roughly -9\,K from the SPT in Fig.~\ref{fig:phasetransition}). From this data, we can see that the heating rate of 1.1\,mK/s at 536\,eV resulted in the highest value of $\bar n$ of 1.23\,$\pm$\,0.02. As expected, the lower the heating rate, the lower $\bar n$ since lower perturbation energy is applied on the surface of the MgO; hence the electron diffraction provides an excellent means of unambiguously mapping the average motional quantum number $\bar n$. The most striking result to emerge from the data is that when the energy of the incident electrons is increased, we observed that $\bar n$ also rises. The latter could be explained by the fact that the key piece of information is a certain cooperative interaction between electron and phonons since many water ice molecules at the surface are very weakly bound, and these will potentially affect as the electron or phonon travels within the lattice. However, remarkably little is known about the characteristic electron--phonon interactions (EPI) during SPT mechanisms; see for example discussions on the cooperative effect of EPI in superconducting transition temperature~\cite{Song2019, Lee2014}. Further studies, which take these variables into account, will need to be undertaken.   


\subsection*{Resolving the strain generated by the Structural Phase Transition in interstellar ice}

From an atomic point of view, the surface atoms of the MgO experience different forces from the atoms within the crystal bulk due to the non--isotropic crystal structural environment. These forces can change the atomic lattice of the surface because the stress generated by the interactions between the atoms and the surrounding elements. Different research suggests the use of theoretical models like Density--Functional Theory and experimental devices such as quartz crystal $\mu$-balance accompanied by experimental techniques like X--Ray Diffraction (XRD) or LEED for the study of this exclusive interaction~\cite{23doc-wlodarczyk2011structures,guevara2018energy, guevara2016detection}. For our case study, we determine the surface stress of the ice by the application of the Williamson--Hall model (W--H). This model approximate the isotropic lattice strain in the MgO lattice, which is then used to estimate the lattice stress by analizing the coherent elastic scattering of electron diffraction. We explore isotropic stress generated by the water monolayer dynamics, which acts until a depth of about (79\,$\pm$\,11)\,nm in the structure of the MgO~\cite{22doc-pottshandbook}. The average penetration depth of the LEED electrons in the material was approximated from the expression developed by Potts~\cite{22doc-pottshandbook} as a function of the electron acceleration voltage and the density of the material. 

The W--H analysis is a quantitative approach, which differentiates between size induced and strain induce peak broadening by considering the peak width as a function of the Bragg diffraction angle ($\theta$) or, in other words, its correspondent value in voltage due to the Bragg's law. This voltage is modulated for our case in the same way that the incident angle is used in the XRD characterization tecnhique. The W--H framework is defined by the following expression: $\beta_{hkl}cos\theta=\frac{k\lambda}{D}+\frac{ 4\sigma sin\theta}{E_{hkl}}$, where $\beta_{hkl}$ corresponds to the broadening of the $hkl$ diffraction peak measured at half of its maximum intensity, $k$ represents the shape factor (assumed to be 0.9), $\lambda$ corresponds to the wavelength of the incident electrons, $D$ denotes the volume--weighted crystallite size, $E_{hkl}$ is the Young's modulus for a specific direction of the crystal planes; and finally $\sigma$ = $E\epsilon$, where $E$ is the isotropic Young's modulus and $\varepsilon$ corresponds to the lattice strain. 

We consider two different experimental heating rates of 0.55\,mK/s and 1.1\,mK/s in order to achieve a fine control of the absorption--desorption processess in comparison with other experiments~\cite{jenniskens1994structural} and additionally, to promote small changes at each time evolution of the SPT in the interstellar ice. Moreover, the fine control of the temperature allows generating small variations in the structural stress of the MgO surface that can be carefully monitored. The use of MgO allows us to understand the stress generated by the absorption--desorption phenomena by the well--known oxide--water interaction where MgO is particularly used. Interaction of MgO with water monolayers occurs by means of the formation of ionic OH groups. These OH groups generate at the MgO surface structure stresses that can change when the STP is generated in the ice structure. Therefore, this makes MgO an excellent mechanical sensor for the laboratory study of the behaviour of interstellar ice.

\begin{figure}[ht]
\includegraphics[width=1.0\textwidth]{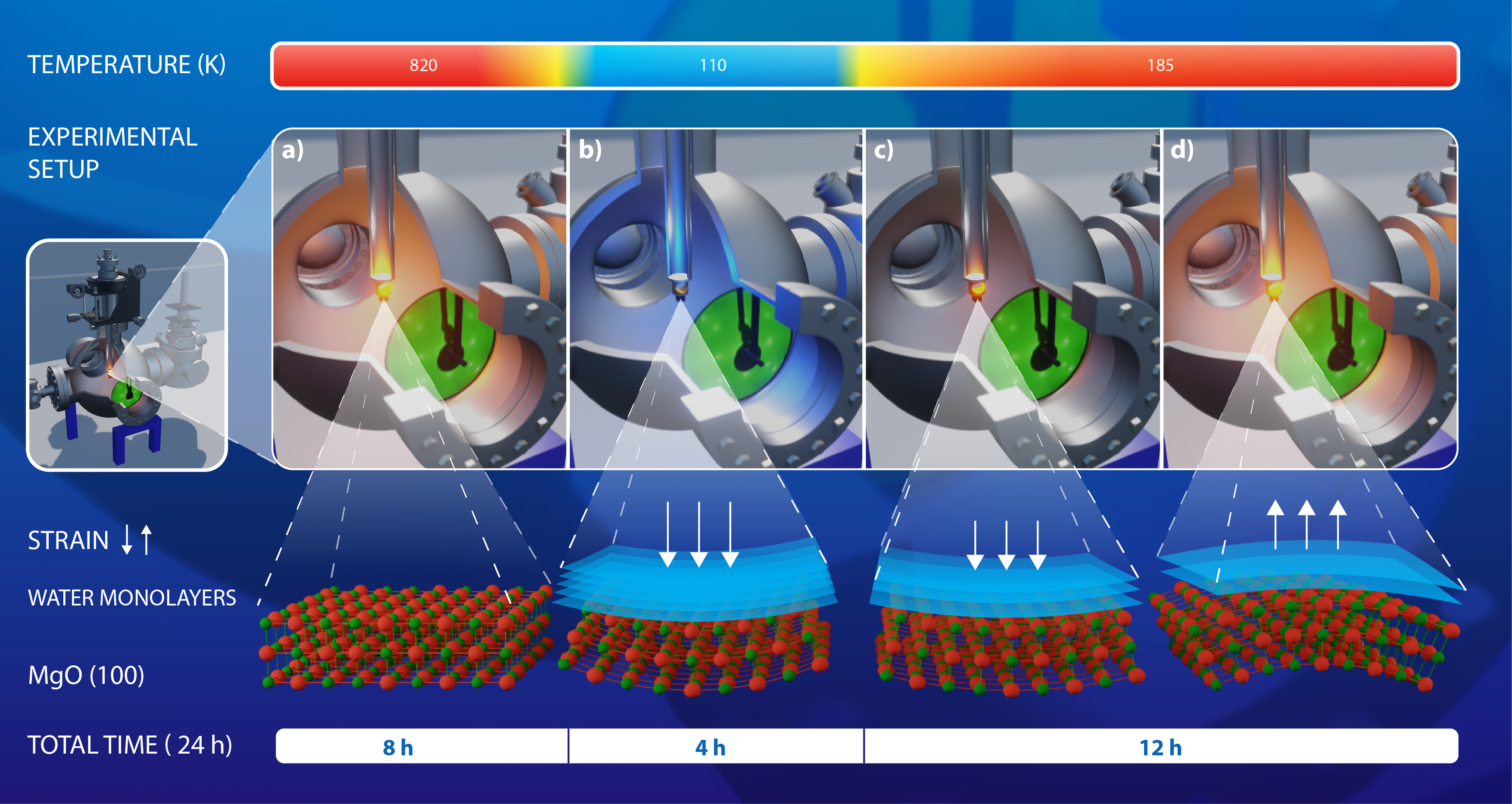}
\caption{\label{fig:Artistic} \textbf{Artistic visualization of the experimental apparatus along with the complete time sequence}. The temperature used in each step is indicated in the colour bar at the top of the experimental setup. The white arrows highlight the magnitude and direction of strain over the MgO surface for the experiments presented in this manuscript. Our experiments in ultra--high vacumm are carried out as follows: \textbf{a)} we began with an annealing process at 820\,K for 8 hours to liberate the residual strains of the MgO surface. Next, \textbf{b)} we performed a cooling process during 4 hours at 110\,K in order to promote the formation of water monolayers, which; therefore, generate compressive strains over the MgO surface. Note the curvature of the MgO crystal structure and also the magnitude of the white strain arrows. Following, \textbf{c)} a low heating rate is employed for 12 hours in order to observe the progressive \textit{c}(4$\times$2) $\longleftrightarrow$ \textit{p}(3$\times$2) transition of the ice at 185\,K. Finally, \textbf{d)} during this period of time, a change of compression to tension along with the Structural Phase Transition is observed by means of the microcantilever MgO sensor. The latter is artistically represented by the change in the direction of the white arrows, the curvature of the MgO crystal structure together with the water monolayers.}
\end{figure}

\begin{figure}[ht]
\includegraphics[width=1.0\textwidth]{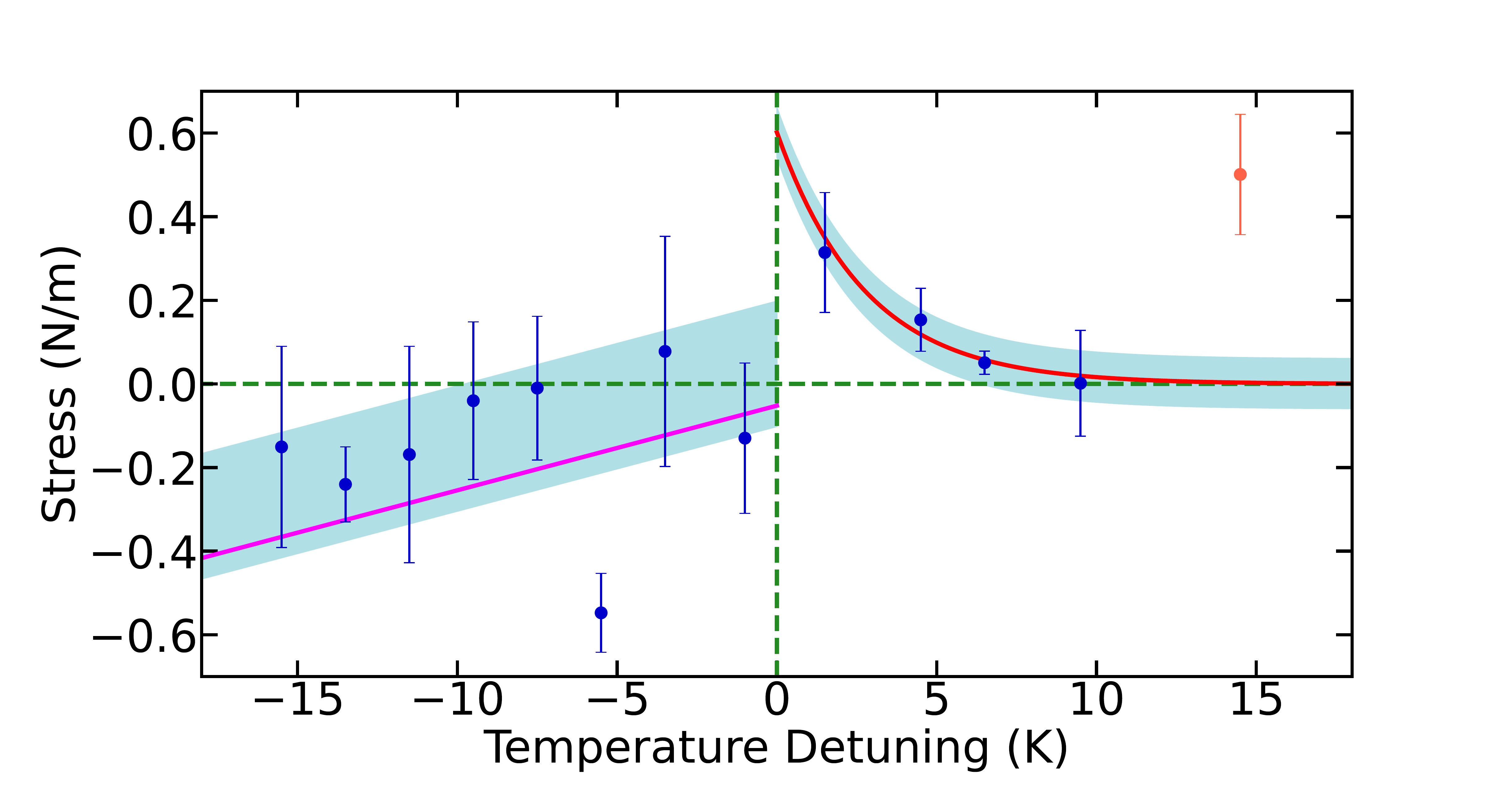}
\caption{\label{fig:StressvsTempDet}\textbf{Surface strain analysis in the transformation of the Structural Phase Transition in interstellar ice.} The experimental data were obtained with the same method as the procedure reported to collect the data presented in Fig.~\ref{fig:phasetransition}. The \textit{c}(4$\times$2) $\longleftrightarrow$ \textit{p}(3$\times$2) transition of ice at 185\,K is used as a reference temperature point (Temperature detuning equals to 0\,K). Each data point (blue dots) corresponds to the results obtained using the W--H stress model. Direct correlation between the Structural Phase Transition (SPT) and the surface strain change from compressive (negative stress) to tensile (positive stress) at 0\,K is noticeably observed. The purple line and blue light area before the 0 K point (left side of the figure) correspond to the linear fit and its standard deviation for the obtained data set. The linear fit represents the decrease tendency of the compressive stress on the MgO lattice. In addition, a little oscillation is observed in the obtained data due to the natural variation of the forces in the MgO lattice when the temperature is increase until the 0 K point. For a guide to the eye, the horizontal dashed--line in green corresponding to zero stress was used. At 0 K, the data set show a change of compressive to tensile stress on the MgO lattice produced by the SPT of the water ice. After 0 K, the process of mechanical stabilization in the MgO lattice surface is shown by the exponential decay fit (red line on the right side of the figure). Finally, the last data point (orange point) outside of the exponential decay, corresponds to an abrupt increase in the tensile stress of the MgO lattice when all the monolayers of water have been released at the last state of our experiment.}
\end{figure}

In our experiment, we began considering three factors that influence the change in adsorption--induced stress calculated for the MgO surface. First, we consider the formation of \textit{n} water monolayers on the MgO surfaces at an initial temperature T$_{1}$ of 110\,K during 4 hours, as shown in Fig.~\ref{fig:Artistic}. The formed water monolayers consist of a structure of ice with symmetry \textit{c}(4$\times$2) that generates normal compressive stress on MgO surface at low temperature. Next, the monolayer structure becomes a detachment process when the system temperature is slowly increased to T$_{2}$ (T$_{1}<$ 185\,K $<$T$_{2}$) during 12 hours. The detachment process is progressive to generate a decrease in the value of compressive stress in the MgO surface. This reduction of the compressive stress is illustrated in Fig.~\ref{fig:StressvsTempDet} as a positive slope before the SPT at 0\,K temperature detuning. The decrease of the compressive stress continues just before 0\,K, where our data show an oscillation between compressive to tensile stress. We attribute this behaviour to the coexistence of the \textit{c}(4$\times$2) and \textit{p}(3$\times$2) ice structure. Furthermore, the effect can be visualized as the change of the stress direction represented in Fig.\ref{fig:Artistic}\,c) and d).

Second, an SPT has been reported by previous LEED experiments in temperatures near to 185\,K~\cite{23doc-wlodarczyk2011structures} related to the change between the ice structures with symetries \textit{c}(4$\times$2) and \textit{p}(3$\times$2). From a structural point of view, this SPT generates a density change due to the difference of greater packaging to a lower packaging between the \textit{c}(4$\times$2) and \textit{p}(3$\times$2) states, respectively. The density change affects considerably the distribution of the water molecules over the surface structure of MgO. It is to be expected that this change also affects the formation of the OH group. Therefore, we attribute the change of compression to tensile stress due to less interaction between the \textit{p}(3$\times$2) ice structure and the MgO surface. This reduction is reflected as an increase of the tensile stress in our data set as shown in Fig.~\ref{fig:StressvsTempDet} immediately after the 0\,K detuning point. This behaviour of the MgO surface can be exploited by using MgO as a type of microcantilever sensor for studying STP during absorption--induced phenomena.    

Once the STP occurs, the water monolayers tend to reach their mechanical stabilization at the same time that the detachment process from the MgO surface continues. This mechanical stabilization is evident in the exponential decay of the tensile stress values near to a value of 0\,N/m, as seen in Fig.~\ref{fig:StressvsTempDet} between 0\,K and +10\,K. Finally, as the third aspect, we consider the continuous detachment of the water monolayer from the MgO surface. At this point, a majority ordering of the new \textit{p}(3$\times$2) structure has already been carried out, so the detachment of the water monolayers begins to be the predominant phenomenon. Our data set confirms that the highest detachment of water monolayer occurs at the range of (200--215)\,K in agreement with the range reported by previous experiments~\cite{23doc-wlodarczyk2011structures}. This detachment generated an abrupt increase in the tensile stress of the MgO surface that is showed near +15\,K temperature detuning in Fig.~\ref{fig:StressvsTempDet}. Here, a progressive loss of many of the water monolayers from the MgO surface is reflecting as the final experimental procedure of our sequence, as we expected. This result highlights a framework to engineer especifically interstellar ice monolayers on top of (100)--oriented MgO substrates to be completely monitored by means of coherent elastic scattering of electron diffraction. 


\section*{Conclusion}

This research describes the dynamics of the Structural Phase Transition (SPT) of interstellar ice at 185\,K. Here, we show for the first time that the use of a very precisely experimentally controlled disturbance is highly desirable for the characterization of SPTs in interstellar ice. By monitoring the intensity of the diffraction spot of the Low--Energy Electron Diffraction (LEED) pattern, we obtain the first experimental evidence of the existence of the SPT between the \textit{c}(4$\times$2) and \textit{p}(3$\times$2) ice structure at 185\,K using the MgO crystal surface as a microcantilever sensor. We describe the \textit{c}(4$\times$2) $\longleftrightarrow$ \textit{p}(3$\times$2) as a quantum two--level system and by carefully controlling the temperature, we explore the possible applications as coherent SPT in interstellar ice for quantum computing. As a consequence, we estimate the amount of phonons caused by the SPT of the ice crystal at 185\,K, which reaches a maximum value of  $\bar n$ = 1.23\,$\pm$\,0.02 inferred from the amplitude ratio of the blue and red motional sidebands. Additionally, we consider possible evidence of cooperative interplay between phonons and electrons that can be explored for a future investigation of electron--phonon strong interactions using a MgO microcantilever based on the minute oscillations of phonons in crystals. Moreover, we  bring to light the use of the Williamsom--Hall model via LEED to study the surface strain behaviour of the MgO surface during the \textit{c}(4$\times$2) $\longleftrightarrow$ \textit{p}(3$\times$2) transition of the ice crystal structures. Our results reveal that the variation of the intensity of the LEED diffracted spots can be used to precisely resolve the performance of the structural planes of solids in the same way that the incident angle is used in X--Ray Diffraction studies. Here, we explore the use of LEED to characterize the compressive to tensile strain change during the SPT in the MgO surfaces. Particularly, we observe that this transition occurs at 185\,K generating a response in the MgO surface produced by the SPT change of packaging inside the crystal considering, therefore, the MgO platform as a novel microcantilever sensor. For future investigations, it might be possible to use a system as the ultrafast LEED described in Ref.~\cite{Horstmann2020} in which the possible existence of qubits in the SPT can be elucidated and likewise explore whether there is magnetic memory in these two structural phases.



\section*{Methods }
\label{Methodology}

Our experimental setup is based on the apparatus used for previous works~\cite{guevara2016detection} and it was maintained at pressures of 10$^{-9}$\,Torr during the experimental results reported here. The LEED patterns were obtained in real time with a commercial ErLEED 1000--A from the SPECS company, and the ranges of the experimental parameters were similar to those used in previous research~\cite{guevara2016detection}. The impurities had been monitored by an Ametek LC-Series quadrupole mass spectrometer coupled to our vacuum system that shows a low level of atmospheric impurities cosisting of moeluclar nitrogen and oxygen, as well as the water characteristic ions in the ultra-hight vacuum system. Figure~\ref{fig:Artistic} provides an artistic visualization of the apparatus used and the time scale of the experiment. The preparation of the experiment began by performing annealing at 820\,K for 8 hours to reverse the formation of ice in the MgO crystal surface and release the stress that may be present in its structure~\cite{guevara2016detection}. Next, a cooling process of 4 hours was carried out to promote the formation of a predominantly ice with \textit{c}(4$\times$2) structure monolayers on the surface of MgO crystal. This process stresses the MgO surface by the formation of ionic OH groups due to the deposition and absorption of free water molecules. Then, to monitor the mechanical behaviour of the MgO surface, we measured the intensity of the LEED diffraction pattern during each experimental cycle. Furthermore, to obtain intensity vs. energy (I--E) characteristic curves, we performed an energy scan with a range of 50\,eV centered around: 311\,eV, 429\,eV, and 536\,eV, and automatically controlled by a LabView script.

To induce the gradual ice \textit{c}(4$\times$2) $\longrightarrow$ \textit{p}(3$\times$2) Structural Phase Transition (SPT), and acquire a better resolution of the measurements, we applied heating rates of 0.55\,mK/s and 1.1\,mK/s. These two rates allowed us to gain insight into the possible data variability caused by the temperature steps taken and the subsequent heating of the water monolayers. We performed measurements every 0.5\,K, and 1.0\,K for the different LEED energies during 12 hours, as shown in Fig.~\ref{fig:Artistic}. From the I--E obtained curves, we estimated the Full With at Half Maximum (FWHM) by a Lorentzian fit and finally use this information along with the Williamson--Hall (W--H) model in order to assess the surface stress induced by the \textit{c}(4$\times$2) $\longrightarrow$ \textit{p}(3$\times$2) SPT. To this end, we were able to relate the energy of the incident LEED electrons to a specific diffraction angle of the Bragg law for each diffraction plane of MgO surface. Thus, assuming that the deformation of the crystal will cause a broadening of the FWHM in the I--E curves, the W--H model can be easily applied. The mean phonon number $\bar n$ was carried out simply by obtaining the ratio of the amplitude motional sidebands (blue sideband/red sideband) for each single experiment via the Lorentzian fit described previously, see Fig.~\ref{fig:phasetransition}.


\bibliography{sample}


\section*{Acknowledgements}

We would like to thank Milena Guevara--Bertsch and Anthony Segura--Garc\'{i}a, for their work and contributions to the first steps of this experiment. We give special thanks to Felipe Molina--Guti\'{e}rrez for his collaboration in the making of the artistic visualization Fig.~\ref{fig:Artistic}. In addition, we would like to thank the anonymous reviewers of this article for their comments on our work; they helped us to improve this manuscript. The authors are very grateful for the support given by the Vicerrector\'{i}a de Investigaci\'{o}n of the Universidad de Costa Rica to carry out this research work. 

\section*{Author contributions statement}

A.C.S, S.M.S, and E.B.J designed the methodology and carried out the experimental measurements. A.C.S, S.M.S, E.B.J, and M.H.C performed the data analysis. A.C.S, S.M.S, and M.H.C prepared the final version of Fig.~\ref{fig:phasetransition}, Fig~\ref{fig:phononnumber}, and Fig~\ref{fig:StressvsTempDet}. O.A.H.S conceived and led the project, was involved in the experimental design, data analysis and interpretation. A.C.S, S.M.S, E.B.J, M.H.C, and O.A.H.S wrote the paper, and all co--authors discussed the results and commend the manuscript. All authors have given approval to the final version of the manuscript.

\section*{Competing Interests statement}

The authors declare that they have no known competing financial interests or personal relationships that could have appeared to influence the work reported in this paper.

\section*{Additional information}



\end{document}